\documentclass[aps,pra,superscriptaddress,amsfonts,amssymb,amsmath]{revtex4}

\usepackage{graphicx}

\newcommand{\be}{\begin{eqnarray}}
\newcommand{\ee}{\end{eqnarray}}

\newcommand{\n}{\nonumber}

\begin{document}


\title{A model of interacting multiple choices of continuous opinions }

\author{C.-I. Chou}
\affiliation{{Department of Physics, Chinese Culture University, Taipei 111, Taiwan, China }}

\author{C.-L. Ho
}
\affiliation
{Department of Physics, Tamkang University, Tamsui 251, Taiwan, China}

\date{Jan 11, 2016}
\begin{abstract}

We present a model of interacting multiple choices of opinions. At each step of the process, a listener is persuaded by his/her neighbour, the lobbyist,  to modify his/her opinion on two different choices of event. Whether or not the listener will be convinced by the lobbyist  depends on the difference between his/her opinion with that of the lobbyist, and with that of the revealed social opinion (the social pressure). If the listener is convinced, he/she will modify his/her opinion and update his/her revealed preference, and proceed to persuade his/her next neighbour. If the listener is not convinced by the lobbyist, he/she will retain his/her revealed preference, and try to persuade the lobbyist to change his/her opinion. In this case, the direction of opinion propagation is reversed. A consensus is reached when all the revealed preference is the same.
Our numerical results show that consensus can always be attained in this model. However, the time needed to achieve consensus, or the so-called convergence time, is longer if the listener is more concerned with the public opinion, or is less likely to be influenced by the lobbyist.

\end{abstract}


\keywords{opinion dynamics, continuous opinions, multiple choices}

 \maketitle   




\section{Introduction}

 In many daily situations it is necessary for a group of people to reach shared decisions or opinions. More often than not, people have different opinions. For instance, whom to vote for before an election; which restaurant to go for dinner, etc.
Despite this,  quite often spontaneous agreement (consensus/majority opinion) can later be reached through discussions/debates etc.
     
The emerging subject of opinion dynamics is concerned with setting up dynamical models to understand how such consensus emerges, or opinion formation. 
One regards the society as a system in which consensus may result from repeated local interactions among the individuals (the agents).
Opinion in mind of any individual has to be quantified by a variable $c$, say, in order for any model building.  Depending on whether $c$ is discrete or continuous, one has model of discrete or continuous opinions formation. Discrete opinions can be binary, with $c=0,1$, representing situations which call for decisions such as yes/no, believe/disbelieve, vote for/against, etc.  It can also be multiple-valued ($c=1,2, \ldots, q$), representing situations in which the degree of preference of an opinion can have $q$ different levels.  The continuum limits of the latter
lead to models with continuous opinions, say with $0\leq c \leq 1$ after appropriate scaling.  This is the case, for instance, when one is asked to rate a book or a movie, ... etc.

 In the past years physicists have attempted to use statistical physics as a framework to study collective phenomena emerging from the interactions of individuals as elementary units in social structures.  Thus Ising models (for binary opinions), $q$-state Potts models (for multi-valued opinions),  Monte-Carlo method, ... etc have been invoked for constructing models of opinion formation (for a review on the subject, see e.g., \cite{CFL,SC}). A general discussion on collective phenomena in social psychology can be found in \cite{G1}.

One of  the oldest models of discrete opinions which mimics binary opinions systems is the voter model \cite{voter} that represents a society in which each agent always follows the opinion of one of his/her nearest neighbours.  The Sznajd model describes a society in which agents are influenced not by any individuals but by groups \cite{Sznajd}.  The majority-rule model mimics a society in which the majority opinion of a randomly chosen group within the total population is assigned to all the agents of that group \cite{G2,G3,G4}.  The Society impact theory describes how individual feel the presence of their peer and how they in turn influence other individuals \cite{SIT}.

The most well-known models of  continuous opinions are the so-called Bounded confidence models. These models consider the realistic aspect of human communication that, while any agent can talk to every other agent, a real discussion  exists only if the opinions of the people involved are sufficiently close to each other.  The two most popular such models are the Deffuant model \cite{D} and the Hegselmann-Krause model \cite{HK}.
In the Deffuant model an agent interacts  only with his/her nearest neighbours, whilst in the Hegselmann-Krause model an agent simultaneously interacts with all other agents whose opinions are within certain prescribed bound. 
More recently, a random  kinetic-exchange type model of continuous opinions is proposed in \cite{LCCC}. Multi-dimensional, or vectorial extensions of the bounded confidence models have also been considered in \cite{FLPR,L}.

As far as we know, all the models proposed so far were for a single choice of opinion, be it discrete (binary or multuple-valued) or continuous. However, in real life, there are situations in which one is faced with multiple choices of opinions. For instance, whether to eat (A) an apple or (B) an orange; to make a choice between (A) hiking in the mountains or (B) having a picnic by the seaside. One can assign binary or continuous values to both opinions (A: apple/mountain; B: orange/seaside). 

In this paper, we would like to propose a model of interacting multiple choices of opinions. As the discrete case can be considered a special case of the continuous one, we shall present the model only for the case of multiple choices of continuous opinions.  We have in mind the situation best illustrated as follows.

Consider a group of kindergarten children lining up in a circle. They are given two choices to decide for the day's activity: Choice (A): hiking in the mountains, and Choice (B): picnic by the seaside. At each step everyone of them will reveal their most preferred choices by showing a card, say ``Red" card for the choice A and ``Blue" card for the choice (B).  So each child knows the majority preference at any time.  This awareness of others' choices is the so-called social pressure.  The process of opinion propagation starts by randomly choosing a child as the first lobbyist to persuade his/her nearest neighbour, the listener. If the listener is convinced, then he/she will enhance the weight of the choice preferred by the lobbyist, and reduce the weight of the other choice according to certain rule. He/she then updates the colour of the his/her card, and proceed to persuade his/her nearest neighbour (in the original direction of opinion propagation). If the listener is not convinced, then he/she will retain the colour card, and try to persuade the lobbyist to adopt his/her preferred choice.  In this latter case, the direction of persuasion is reversed.  And this process goes on until a stationary state is attained. 
The stationary state may be one that consists of divided opinions among the agents, or one with consensus on one of the two choices.

This paper is organised as follows.  In Sect.\,II we define the model.  Sect.\,III presents numerical results that demonstrate some properties of the model.
Generalisation to models with more than two choices is briefly described in Sect.\,IV.  The last section summarises the paper.


%
%

\section{The model}

We consider a model of interacting multiple continuous opinions as follows.  For simplicity and definiteness, we shall discuss the situation where only two different opinions, $A$ and $B$, need be decided. Generalisation to more opinions is straightforward, and will be mentioned later.

Let there be $N$ agents arranged in a circle. This is just one dimensional periodic boundary conditions. 
At each step of the process every agent has in mind certain degree of preference of the two opinions $A$ and $B$. 
Every agent will reveal his/her preferred choice according to which opinion has the greater weight at that moment. The ratio of the two revealed opinions serves as a reference to every agent in the next step in his/her decision as to whether he/she want to change opinion when persuaded by his/her neighbour. 

Below we describe our model in three steps: 1) set up of  the system, 2) persuasion process, 
and 3) internal transformation of opinion state and propagation of opinion.

\subsection{System set up}

\begin{enumerate}

\item[$\bullet$]

{\bf Representation of opinion state}:

 ~~~The opinion in the mind of any agent $(i=1,\ldots, N)$ at each step $t=0,1,2,\ldots$ is written in matrix form as a two-component state function $\psi_i (t)$ as
\be
 \psi_i (t) =
 \left( \begin{array}{c}
 c_{i,A}(t) \\
 c_{i,B} (t) \end{array}\right).
\end{eqnarray}
Here $ 0<c_{i,A}(t), c_{i,B}(t) <1$ give the measure of the degree of preference  of  $A$ and $ B$ of the $i$-agent at step $t$.


\item[$\bullet$]

{\bf Revealed preference}:

~~~At each step every agent will show his/her preference to the other agents. Which choice of opinion $A$ and $B$  is shown to the others 
is determined by the factor
\be
p_{i, \tau}(t)\equiv \frac{c_{i,\tau}(t)}{c_{i, A}(t)+c_{i,B}(t)},~~\tau=A, B.
\ee
When  $ p_{i, A}(t) > p_{i, B}(t)$, the agent will show the choice $A$, otherwise he/she will show $B$ to the others. 

~~~The revealed preference of opinion $A$ is defined by the ratio of the total number of agents choosing $A$ at step $t$  among the $N$ agents:
\be
p_{s, A}(t)\equiv \sum_{i=1}^N \frac{[p_{i,A}(t)+0.5]}{N},
\ee
where $[x]$ represents the integral part of $x$. 
The corresponding revealed preference of opinion $B$ is simply given by $ p_{s, B}(t)=1-p_{s, A}(t)$. 
These revealed preferences in some sense represent the social pressure to each and every agent in their decision in the next step.

~~~When either $p_{s, A}(t)=1$ or $p_{s, B}(t)=1$, the group of agents reaches a consensus of opinion $A$ or $B$, respectively.

\end{enumerate}

\subsection{Persuasion process}
 
\begin{enumerate} 
 
\item[$\bullet$] {\bf Initial setting}:
 
~~~The initial opinion states of every agent, characterised by the coefficients $c_{i,A}$ and $c_{i,B}$,  will be randomly generated. 
One then randomly selects an agent $j$ and his/her neighbour $i$($i=j+1\: {\rm or}  \:j-1$) as the first lobbyist and the first listener, respectively.

\item[$\bullet$] {\bf Decision factor}:

~~~We shall consider in this paper the situation where the listener is more likely to change his/her mind if his/her preference differs too much from the majority opinion revealed (social pressure/bandwagon effect), and is more likely to be convinced by the lobbyist if their opinions do not differ too much (peer effect). 

~~~The influence of these two factors on the opinion state of the listener can be quantified by a decision factor, $P_{i,\alpha}(t)$, representing the social pressure (gauged by a parameter $0\leq \alpha\leq 1$) and 
the influence of the lobbyist ($j$-agent) on the $i$-person:
\be
P_{i,\alpha}(t)\equiv \alpha |p_{s,A}(t)-p_{i,A(t)}| + \left(1-\alpha\right) \left(1-|p_{j,A}(t)-p_{i,A}(t)|\right).
\label{factor}
\ee
The first term represents the difference between the opinions of the agent and the majority of the group, and the second term gauges the difference between the opinions of  the agent and the persuader. 
$P_{i,\alpha}(t)$ is large when the preference of the agent differs greatly from that of the majority and closes to that of the lobbyist. 

~~~We then compare $P_{i,\alpha}(t)$ with a reference (random) number $ r(t)\in [0,1]$ (accounting for the state of mind/mood of the listener at that time), 
\be
&  P_ {i,\alpha}(t) >r (t) &: {\rm\ persuasion\  successful};\n \\
& P_{i,\alpha}(t)< r(t) & : {\rm\ persuasion\  failed}.
\label{compare}
\ee

\end{enumerate}


\subsection{Transformation of opinion states and Propagation of opinion}

Depending on whether the listener is convinced or not by the lobbyist, he/she will have to modify his/her opinion state and decide to whom he/she should  persuade.   For our model we shall adopt the following rules:

\begin{enumerate}
\item[$\bullet$] 
 $P_{i,\alpha}(t)>r (t)$ : persuasion successful, the $i$-agent will update  his/her opinion state, and proceed to persuade his/her neighbour in the same direction, i.e., the $i+1$-agent away from the original lobbyist.

~~~For simplicity, we assume in this model the opinion state of the $i$-agent is updated according to the influence of the lobbyist, proportional to the  difference in their opinions measured by $( p_{j,A}(t)-p_{i,A}(t))$ and gauged by a parameter $0\leq \mu\leq 1$. :
\be
c_{i,A}(t+1)&=&\left[ c_{i,A(t)}+\mu \left( p_{j,A}(t)-p_{i,A}(t) \right)\right],~~([.]: {\rm integral\  part})\n\\
 c_{i,B}(t+1) &=&\left[ c_{i,B}(t)+\mu \left(p_{j,B}(t)-p_{i,B}(t) \right)\right]\n\\
                      &=&\left[ c_{i,B}(t)-\mu \left( p_{j,A}(t)-p_{i,A}(t) \right)\right],~({\rm note}\  p_B(t)=1-p_A(t))\n
\ee

Taking the integral part is to ensure that the coefficients stay in their defined range, i.e., when the coefficient $c_{i,A}(t)$ (or $c_{i,B}(t)$) $ >1$ (or $<0$), it is set to
$ 1 (0)$.

\item[$\bullet$]  $P_{i,\alpha}(t) <r(t)$ : persuasion failed, the $i$-agent retains opinion state, and proceeds to persuade neighbour in the reversed direction, i.e., the original lobbyist.

 \end{enumerate}
 
Stationary state is reached when the opinion states of these $N$ agents become fixed.
Consensus is defined as the state when all the $N$ agents agree on the same opinion, either $A$ or $B$.

%
%

\section{Numerical results}

We have simulated the model described above for a number of parameters $\alpha, \mu$ and $N$. The main properties under study are whether stationary state, and state with consensus in particular, can be attained, and the average time (number of steps) to attain it. For each set of parameters, a large number of initial opinions is randomly set up, and the ensemble average of the time needed to reach stationary state/consensus state is computed.

In Figs.\,1 and 2, we show the time, or rather the number of steps, required for a system of $N=25$ agents to reach a consensus with different choices of $\alpha$ and $\mu$.

Figs.\,1(a) and 1(b) show the situations in which the initial majority opinion, here the choice B, is enhanced and finally became the final consensus. These figures also show that, with the same parameters $\alpha$ and $\mu$, the time needed to reach a consensus can be very different.

Figs.\,1(c) and 1(d) show that there are situations where consensuses were  reached only after several exchanges of majority opinions,  and that the final consensus could be a different choice with the same set of parameters, e.g.,  the choice B in Fig.\,1(c), and A in  Fig.\,1(d).

Fig.\,2 indicates (compared with the Fig.\,1 with the same $\alpha$) that when $\mu$ becomes smaller,  the system takes longer steps to reach a consensus. This is easily understandable: when the listener cares more about the public opinion
(larger $\alpha$),  he/she is more reluctant to be persuaded by the lobbyist.

In Figs.\,3-5 we show the pictures of  frequencies of the values $(c_A, c_B)$ for various parameters. This gives an idea of how these coefficients evolve. Consensus is reached when all the points are located below (with choice A) or above (with choice B) the line $c_A=c_B$.  Note that in our model consensus is defined by all $c_A>c_B$ or $c_B>c_A$.  Hence it is not necessary, as is evident from the figures, that $c_A=1$ or $c_B=1$ for the all agents in the final state.

To better understand the properties of the model, we have simulated the model with $1000$ random initial set up, and determine the ensemble average of the time needed to reach a consensus with different sets of the parameters $\alpha$ and $\mu$.  These are shown in Figs.\,6-9. 

In Figs.\,6 and 7, we plot the average consensus time $\langle T\rangle$ and $1/\langle T\rangle$ versus $\mu$ for different $\alpha$. It is seen that generally for a fixed $\mu$, $\langle T\rangle$ increases as $\alpha$ increases. In Figs.\,8 and 9, we plot the average consensus time $\langle T\rangle$ and $1/\langle T\rangle$ versus $\alpha$ for different $\mu$. For a fixed $\alpha$, $\langle T\rangle$ increases as $\mu$ decreases. This implies that the listener is harder to be persuaded by the lobbyist, if he/she is more concerned with the public opinion (larger $\alpha$), or if the opinion of the lobbyist has a lesser weight in the listener's mind (smaller $\mu$).

Lastly, in Figs.\,10-11 we present 3-dimensional plots of $\langle T\rangle$ versus $\alpha$ and $\mu$ with an ensemble size of $200$ for each set of parameters.  It is seen that   
$\langle T\rangle$ is higher near $(\alpha, \mu)=(0,0)$ and $(1,0)$.  This is understood as the listener being harder to persuade when he/she is indifferent to the lobbyist's opinion ($\mu\sim 0$), or  he/she is very concerned with the public opinion.  The average consensus time needed also increases as the number of agent $N$ increases.  This is in conformity with the usual experience.

%
%

\section{Generalization}

The above model can be easily generalised to system with multiple choices of opinions, say $q$ opinions.  The main changes are summarised as follows.

 The opinion in the mind of any agent $(i=1,\ldots, N)$ at each step $t=0,1,2,\ldots$ is represented as
$
 \psi_i (t) =\{
 c_{i,1}(t), c_{i,2} (t), \ldots,c_{1,q}(t)\}, ~ 0\leq c_{i,k}(t)  \leq 1, ~~k=1,2,\ldots, q.
 $

~~~At each step every agent will show his/her preference to the other agents according to the maximal value of the factor
\be
p_{i, \tau}(t)\equiv \frac{c_{i,\tau}(t)}{\sum_\tau c_{i, \tau}(t)},~~\tau=1,2,\ldots, q.
\ee

The revealed preference of opinion $\tau (\tau=1,2,\ldots, q)$ is defined by the ratio of the total number of agents choosing $\tau$ at step $t$  among the $N$ agents:
\be
p_{s, \tau}(t)\equiv \frac{1}{N}\sum_{i=1}^N \left[\frac{p_{i,\tau}(t)}{{\rm max}_\sigma\{p_{i,\sigma}\}}\right],~~\tau,~\sigma=1,2,\ldots,q.
\ee
where again $[x]$ represents the integral part of $x$.  
 
 The decision factor, $P_{i,\alpha}(t)$, representing the social pressure (gauged by a parameter $0\leq \alpha\leq 1$) and 
the influence of the lobbyist ($j$-agent) on the $i$-person is given by
\be
P_{i,\alpha}(t)\equiv  \frac{1}{q}\sum_{\sigma=1}^q\left\{\alpha |p_{s,\alpha}(t)-p_{i,\alpha(t)}| + \left(1-\alpha\right) \left(1-|p_{j,\alpha}(t)-p_{i,\alpha}(t)|\right)\right\}. 
\label{factor}
\ee
The first term represents the difference between the opinions of the agent and the majority of the group, and the second term gauges the difference between the opinions of  the agent and the persuader. 
$P_{i,\alpha}(t)$ is large when the preference of the agent differs greatly from that of the majority and closes to that of the lobbyist. 
As before,  one then compares $P_{i,\alpha}(t)$ with a reference (random) number $ r(t)\in [0,1]$ (accounting for the state of mind/mood of the listener at that time), 
\be
&  P_ {i,\alpha}(t) >r (t) &: {\rm\ persuasion\  successful};\n \\
& P_{i,\alpha}(t)< r(t) & : {\rm\ persuasion\  failed}.
\label{compare}
\ee

When persuasion is successful, the opinion state of the $i$-agent is updated according to the influence of the lobbyist by
\be
c_{i,\tau}(t+1)&=&\left[ c_{i,\tau(t)}+\mu \left( p_{j,\tau}(t)-p_{i,\tau}(t) \right)\right], ~~\tau=1,2,\ldots,q.
\ee

Stationary state is reached when the opinion states of these $N$ agents become fixed.
Consensus is defined as the state when all the $N$ agents agree on the a particular opinion $\tau$..

%
%

\section{Summary}

In this work we have presented a model of interacting multiple choices of opinions. At each step of the process, a listener is persuaded by his/her neighbour, the lobbyist,  to modify his/her opinion on two different choices of event. Whether or not the listener will be convinced by the lobbyist  depends on the difference between his/her opinion with that of the lobbyist, and with that of revealed social opinion (the social pressure). If the listener is convinced,  he/she will modify his/her opinion and update his/her revealed preference, and proceeds to persuade his/her next neighbour. If the listener is not convinced by the lobbyist, he/she will retain his/her revealed preference, and try to persuade the lobbyist to change his/her opinion. In this case, the direction of opinion propagation is reversed. A consensus is reached when all the revealed preference is the same.

The cases considered in our numerical simulation show that consensus can always be attained. However, the time needed to achieve consensus, or the so-called convergence time, is longer if the listener is more concerned with the public opinion (as measured by the parameter $\alpha\sim 1$ in our model), or is less likely to be influenced by the lobbyist (as measured by the parameter $\mu\sim 0$ in our model).

The model presented here is bi-directional in that the direction of the propagation of opinion can be forward or backward, depending on wether the listener is convinced or not.
One could also consider a model which is  uni-directional: the listener will try to change his/her next nearest neighbour in the forward direction with his modified/original opinion depending on whether he/she is convinced or not by the previous lobbyist.

A more interesting generalisation of the model is to consider a system with more than two possible choices of opinions, as briefly discussed in Sect.\, IV. One would expect more complicated phase structures in the opinion space, such as polarisation or fragmentation of opinions. Such systems are now under investigation. 
\newpage

\centerline{\bf Acknowledgments}

The work is supported in part by the Ministry of Science and Technology (MoST)
of the Republic of China under Grant NSC-102-2112-M-032-003-MY3.  Preliminary version of this work was presented by CLH at the ``Workshop
of Quantum Simulation and Quantum Walks" held at Yokohama National University during Nov 16-18, 2015. He thanks Y. Ide and Y. Shikano for helpful comments.

\newpage

\begin{figure}[ht] \centering
\includegraphics*[width=8cm,height=8cm]{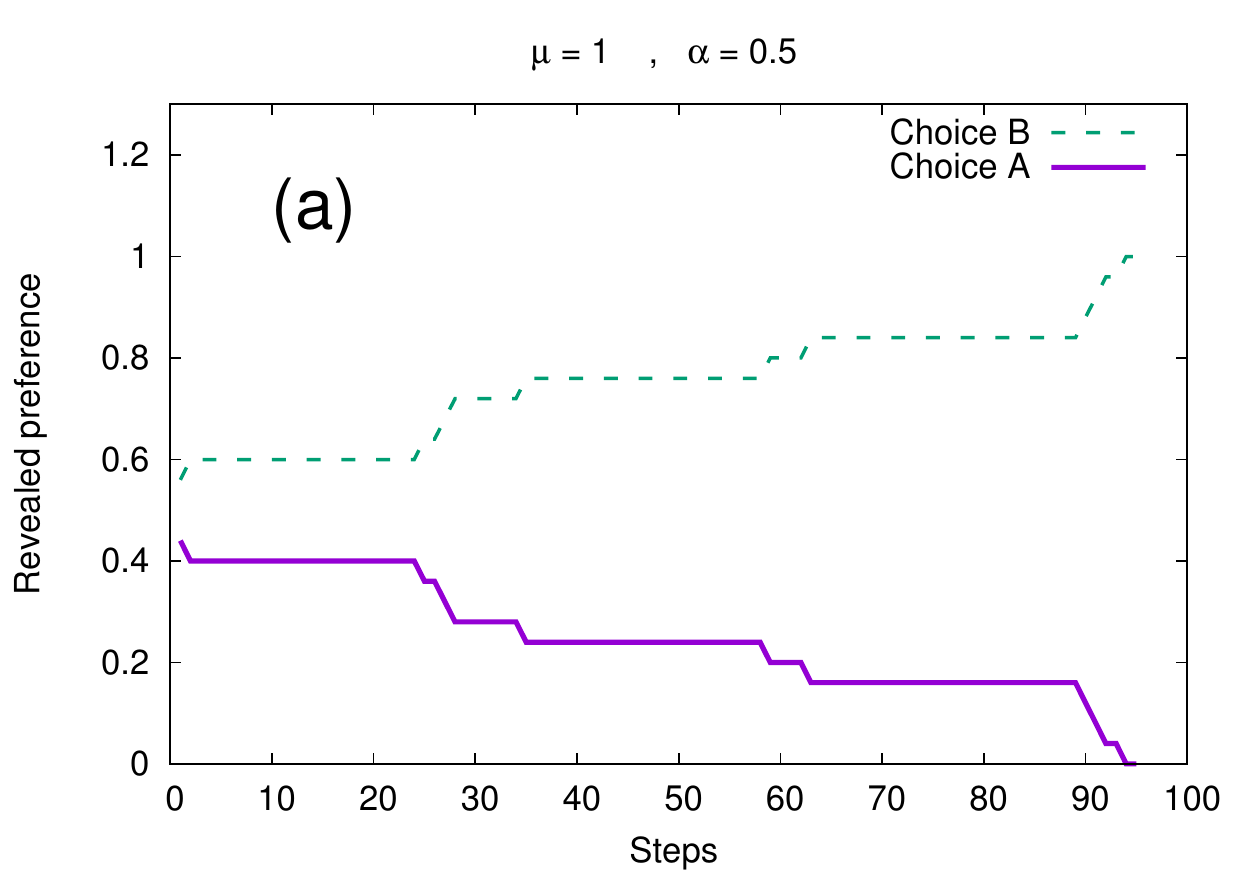}\hspace{1cm}
\includegraphics*[width=8cm,height=8cm]{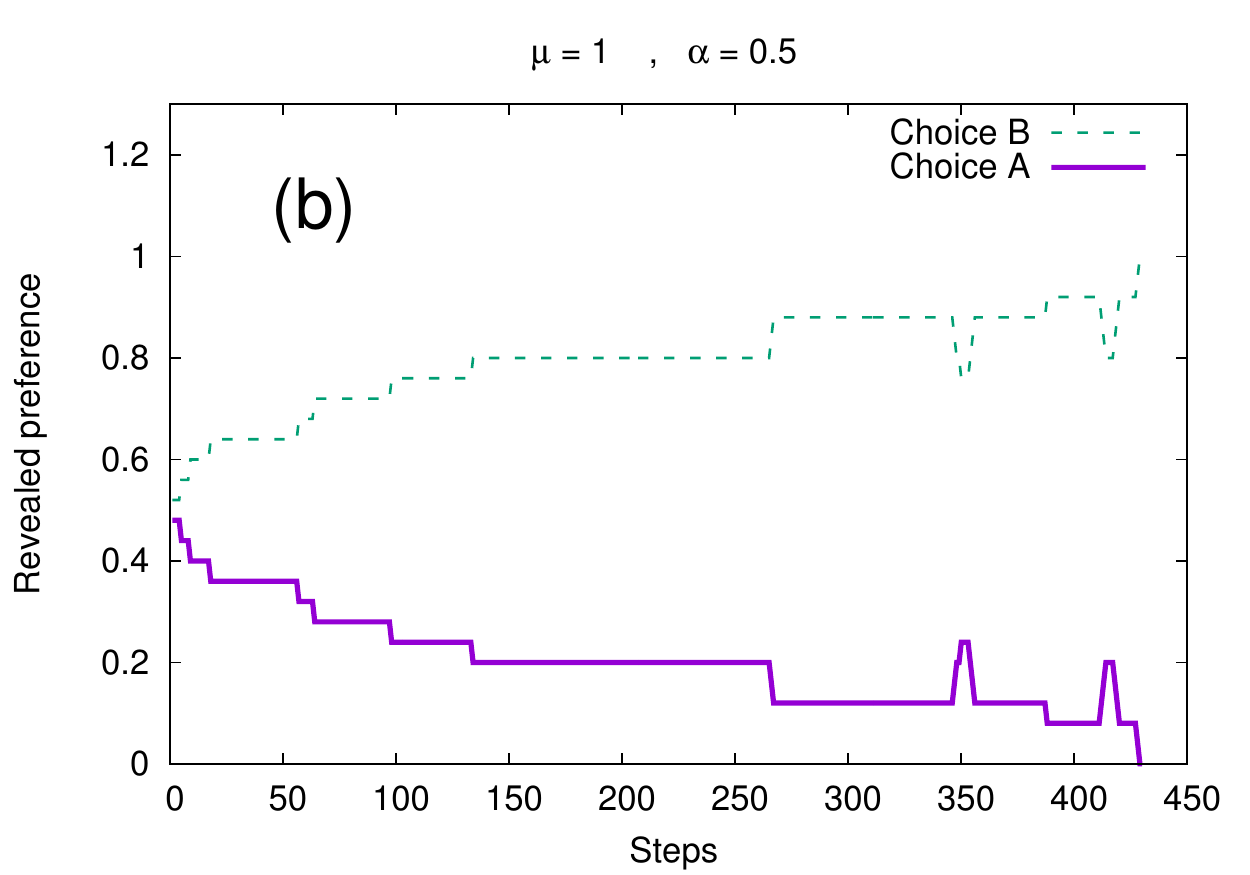}\\
\includegraphics*[width=8cm,height=8cm]{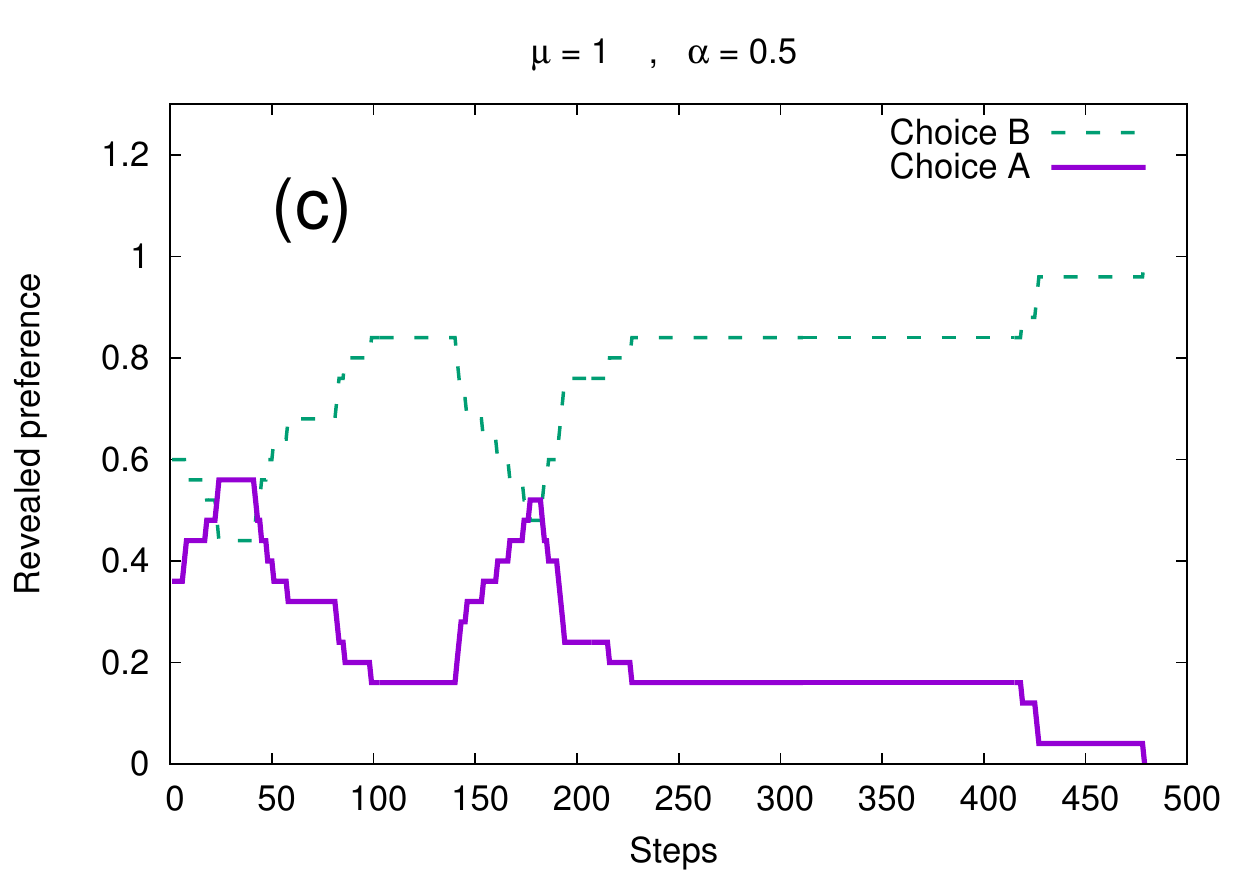}\hspace{1cm}
\includegraphics*[width=8cm,height=8cm]{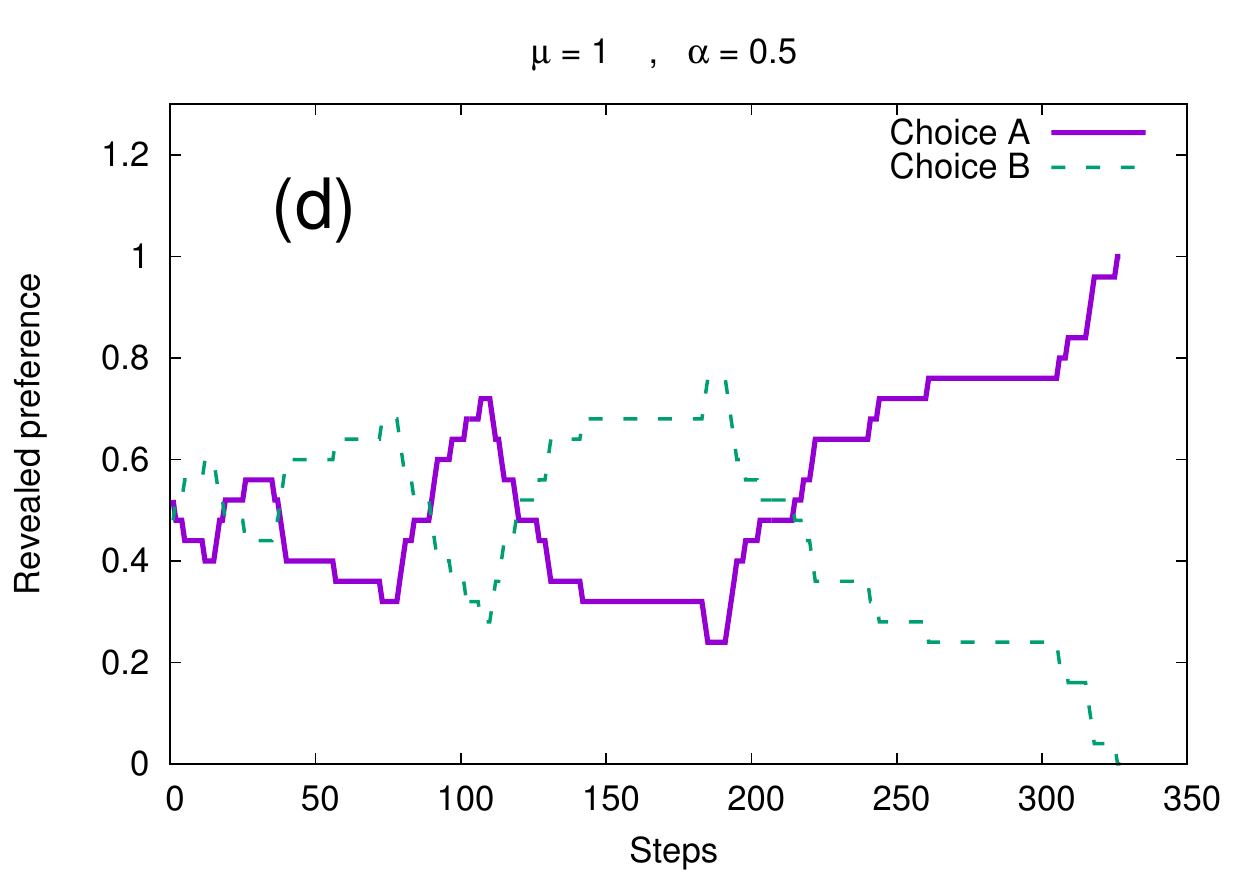}
\caption{Formation of consensus 1: Plot of revealed preference $c_{A,B}$ versus number of steps $T$ 
with $\alpha=0.5$, $\mu=1$.} \label{fig:EX0}
\end{figure}

\begin{figure} 
 \centering
 \includegraphics{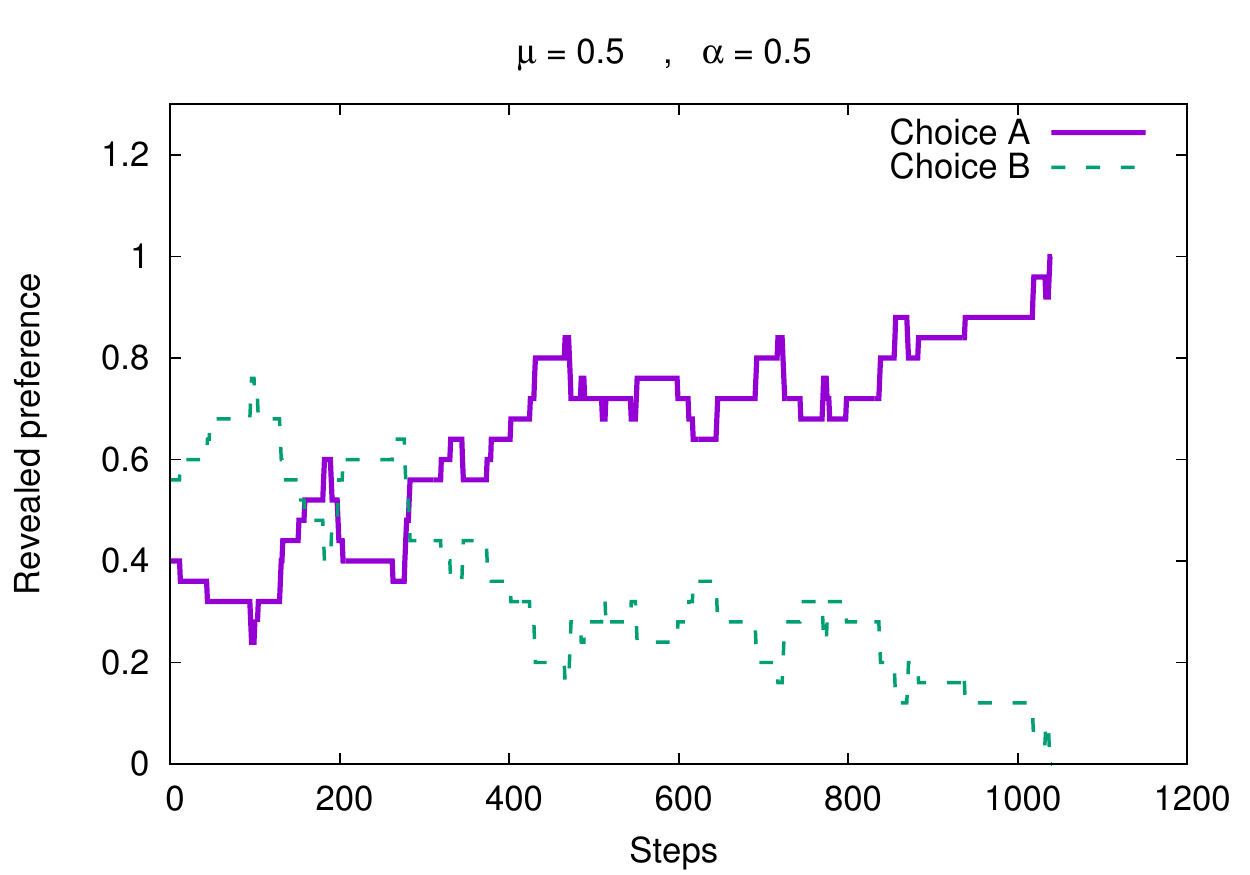}
\caption{Formation of consensus 2: Plot of revealed preference $c_{A,B}$ versus number of steps $T$ 
with $\alpha=0.5$, $\mu=0.5$.}
\label{fig2}
\end{figure}

\begin{figure}[ht] \centering
\includegraphics*[width=8cm,height=8cm]{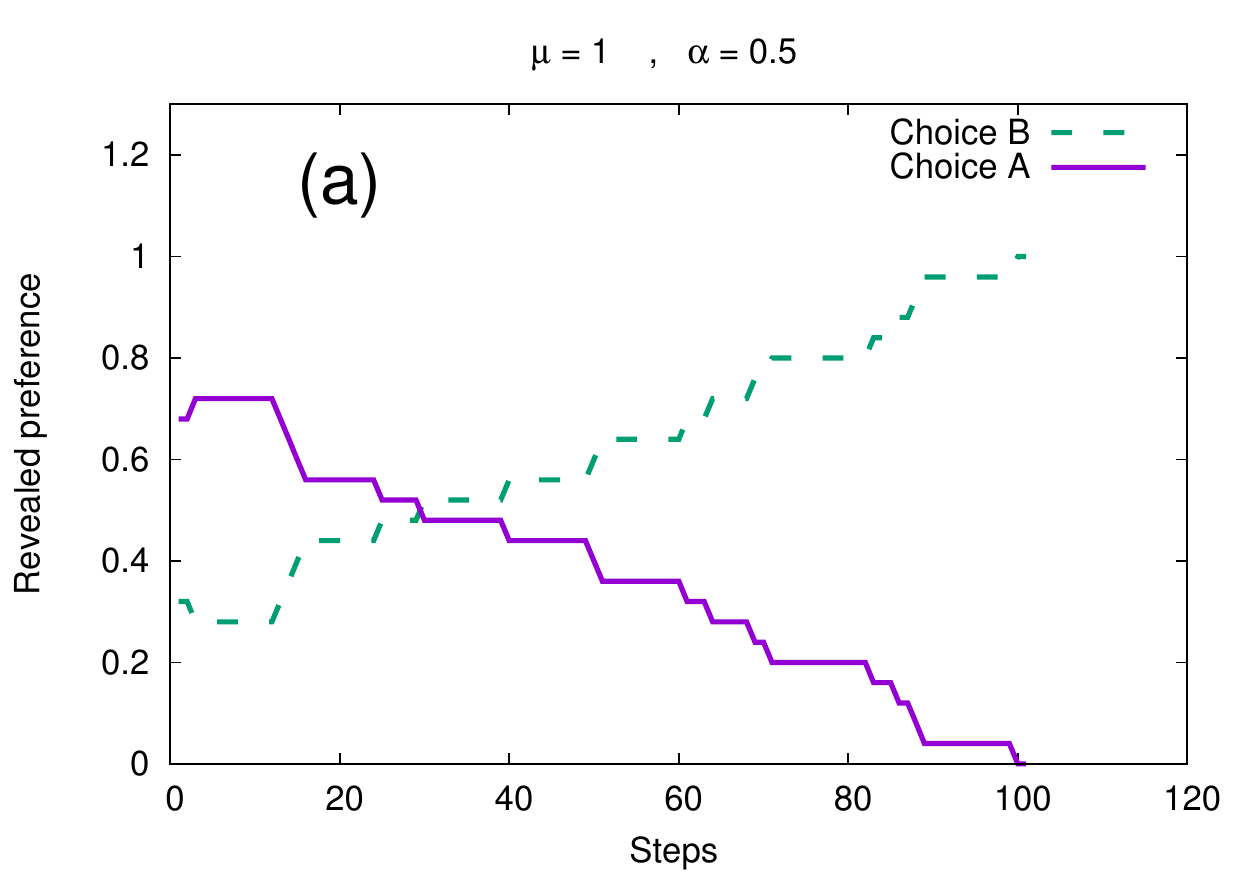}\hspace{1cm}
\includegraphics*[width=8cm,height=8cm]{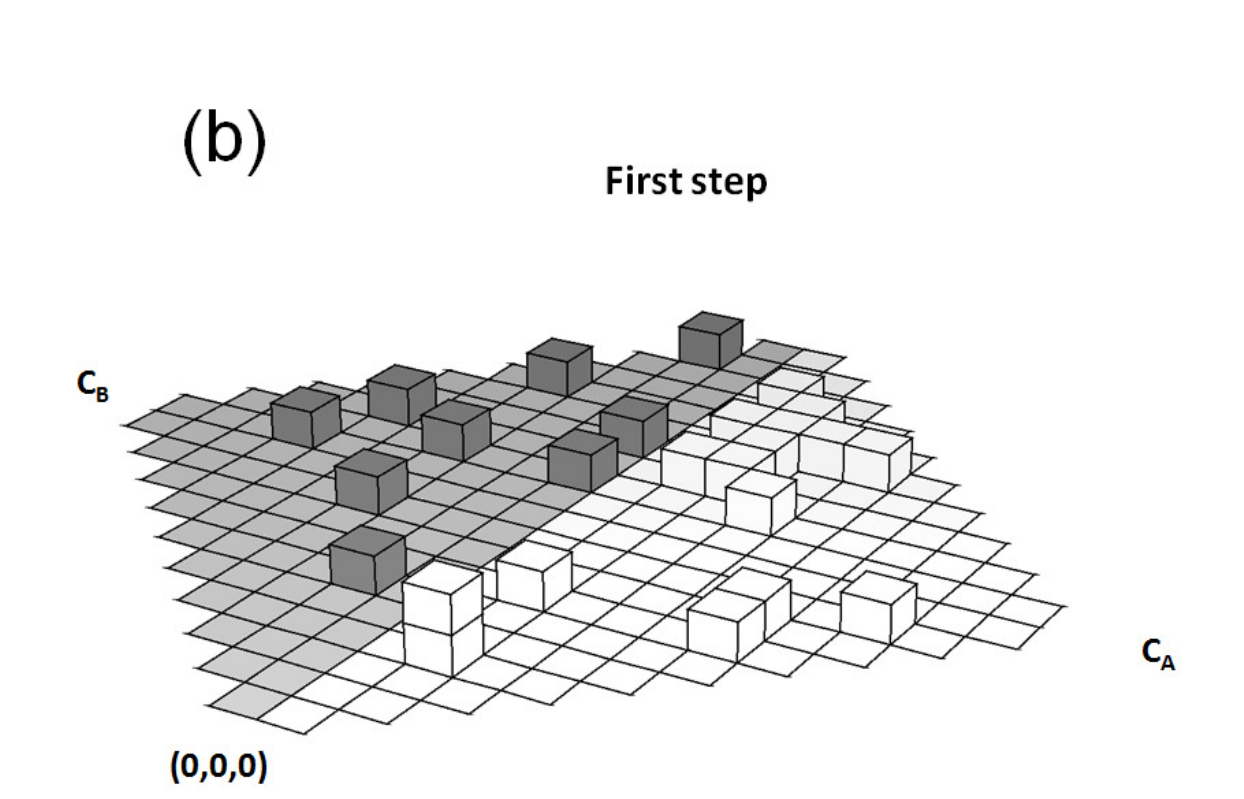}\\
\includegraphics*[width=8cm,height=8cm]{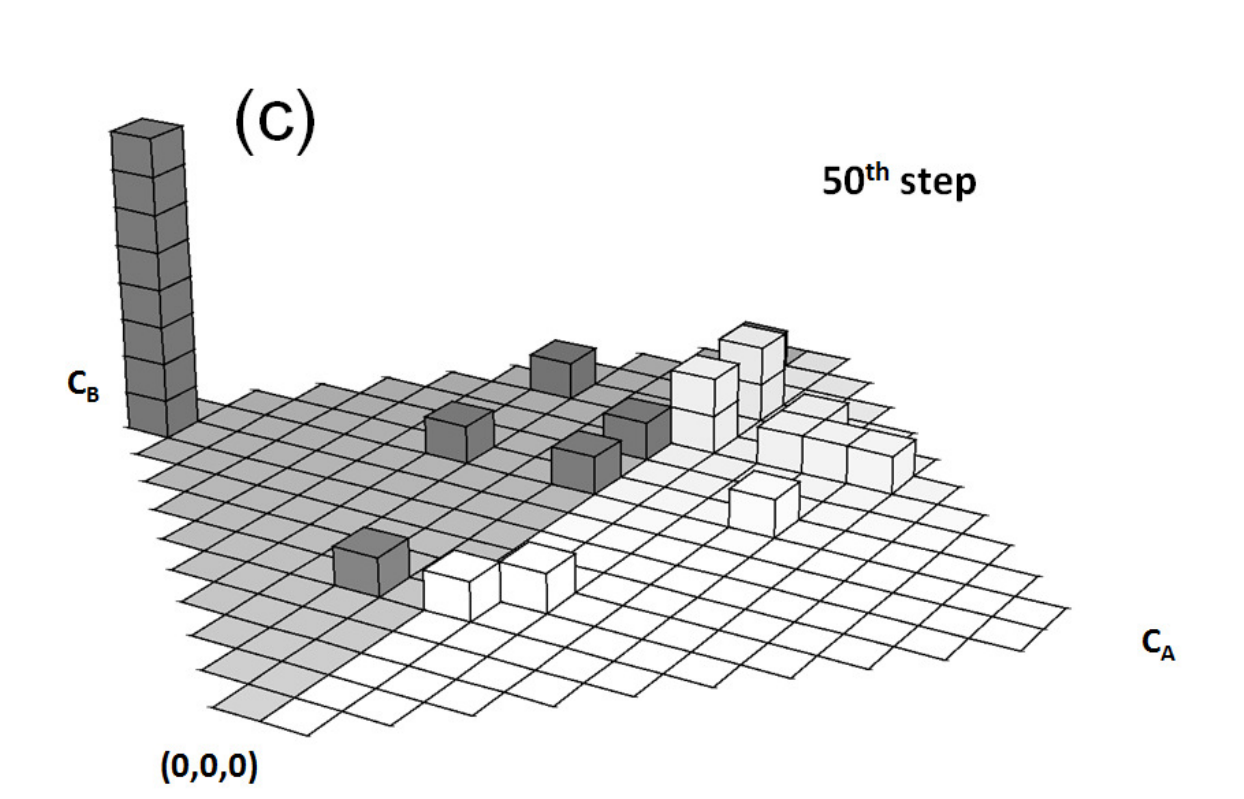}\hspace{1cm}
\includegraphics*[width=8cm,height=8cm]{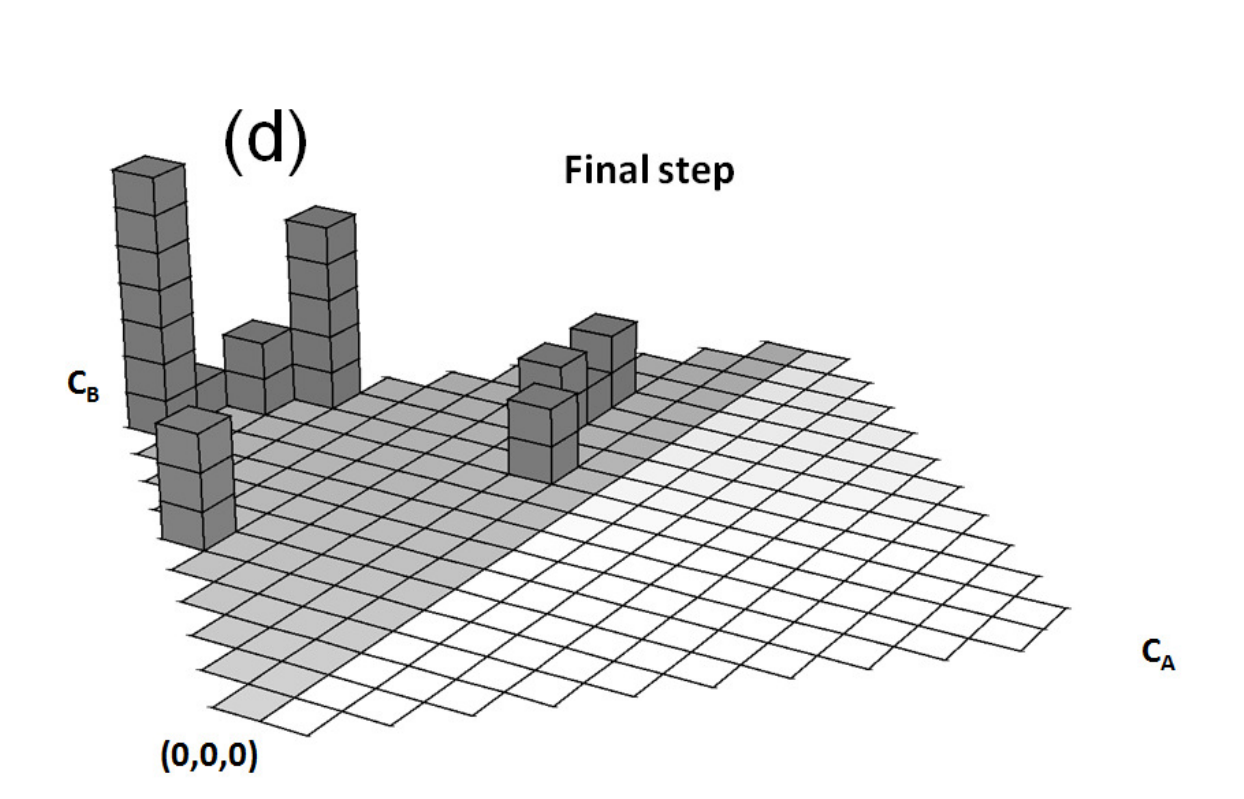}
\caption{Formation of consensus 3:(a) Plot of revealed preference $c_{A,B}$ versus number of steps $T$ 
with $\alpha=0.5$, $\mu=1$, (b)  frequency of $c_{A,B}$ at the first step, (c)  frequency of $c_{A,B}$ at the $50^{th}$ step, (d)  frequency of $c_{A,B}$ at the final step. Consensus reached is choice B.} \label{fig3}
\end{figure}

\begin{figure}[ht] \centering
\includegraphics*[width=8cm,height=8cm]{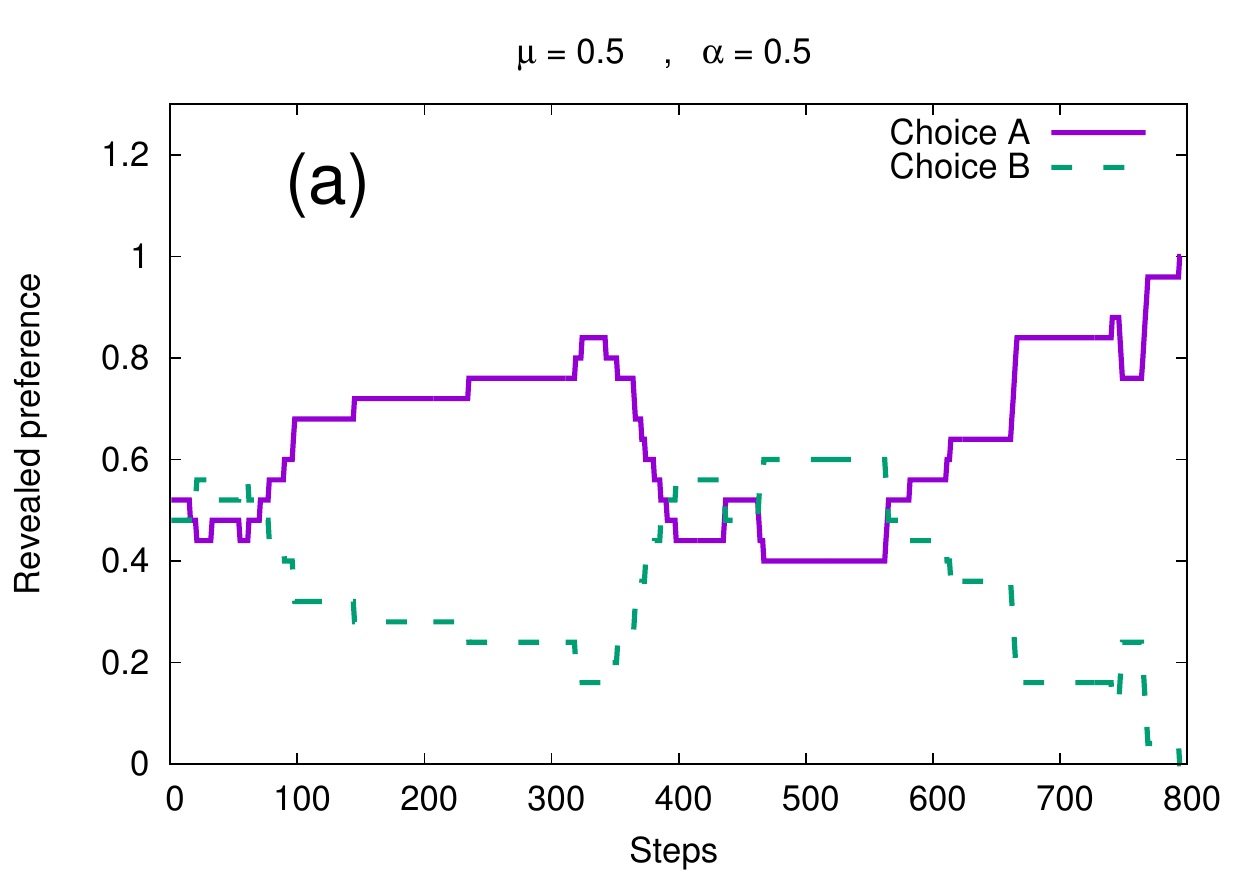}\hspace{1cm}
\includegraphics*[width=8cm,height=8cm]{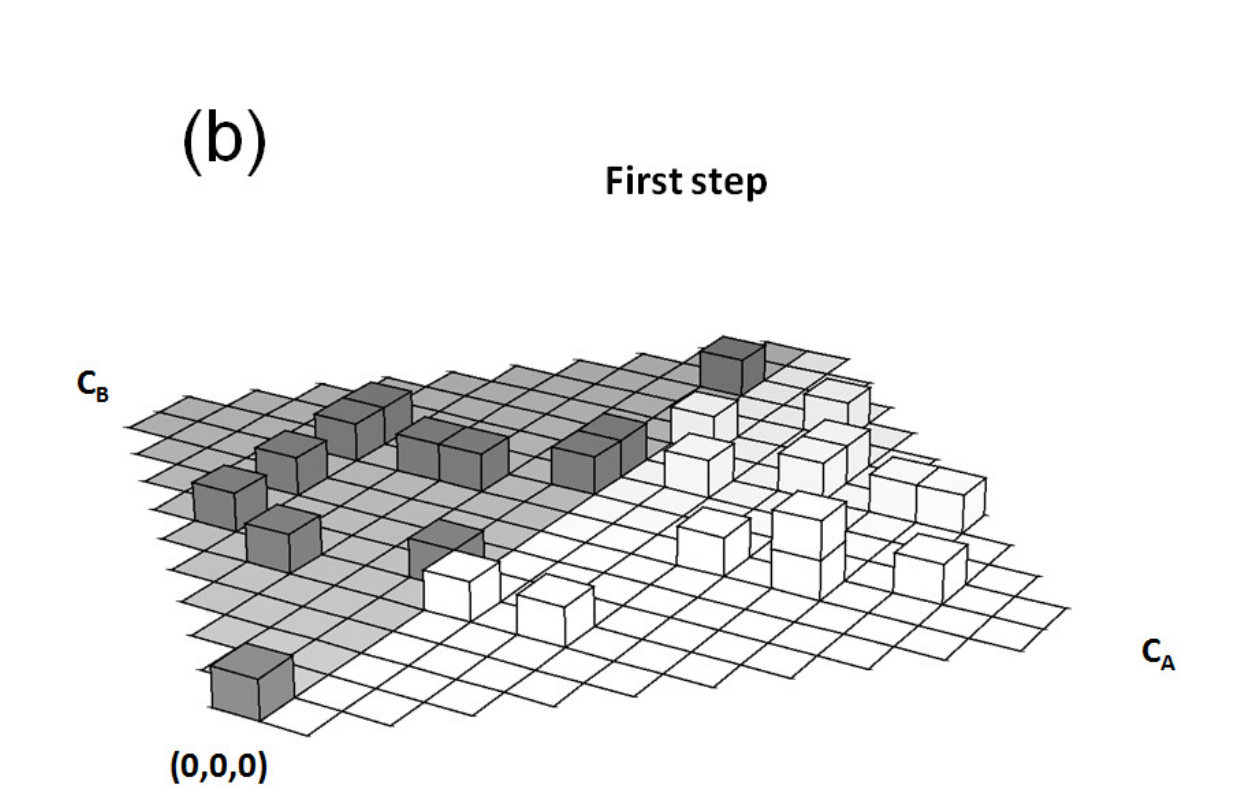}\\
\includegraphics*[width=8cm,height=8cm]{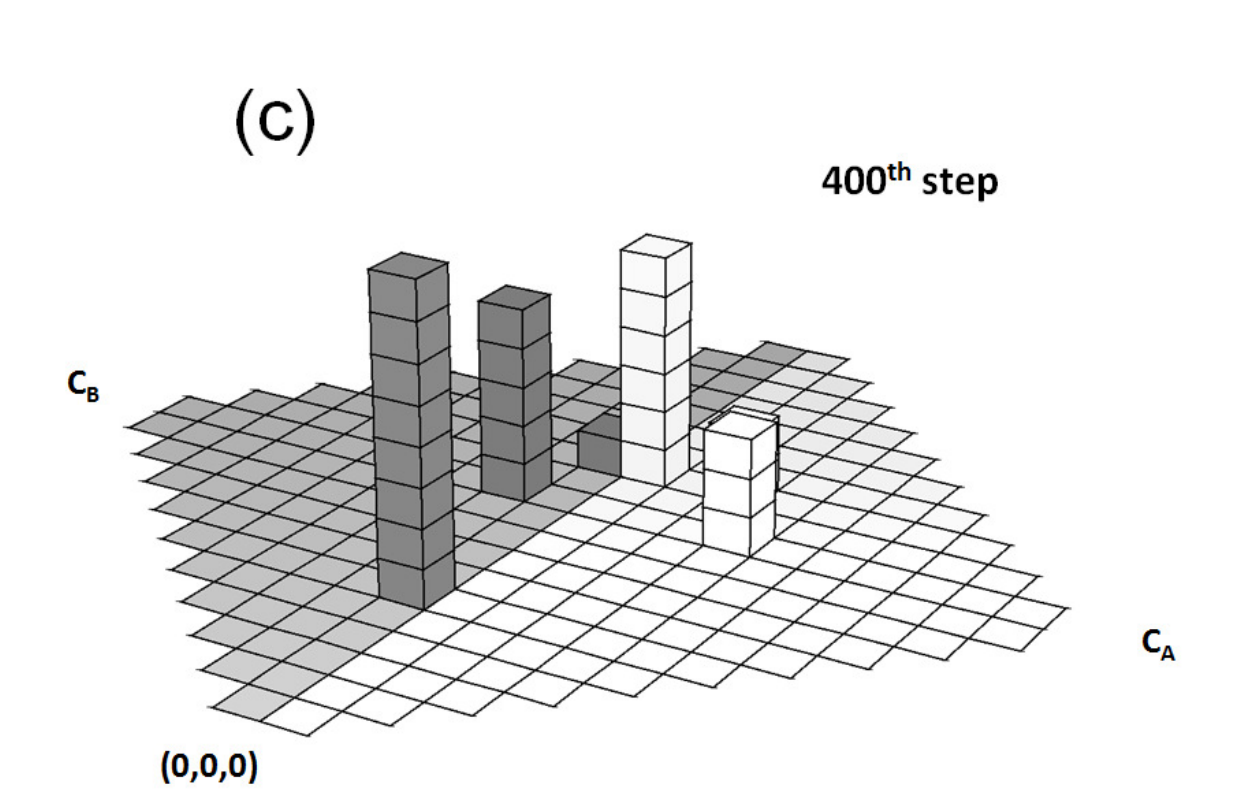}\hspace{1cm}
\includegraphics*[width=8cm,height=8cm]{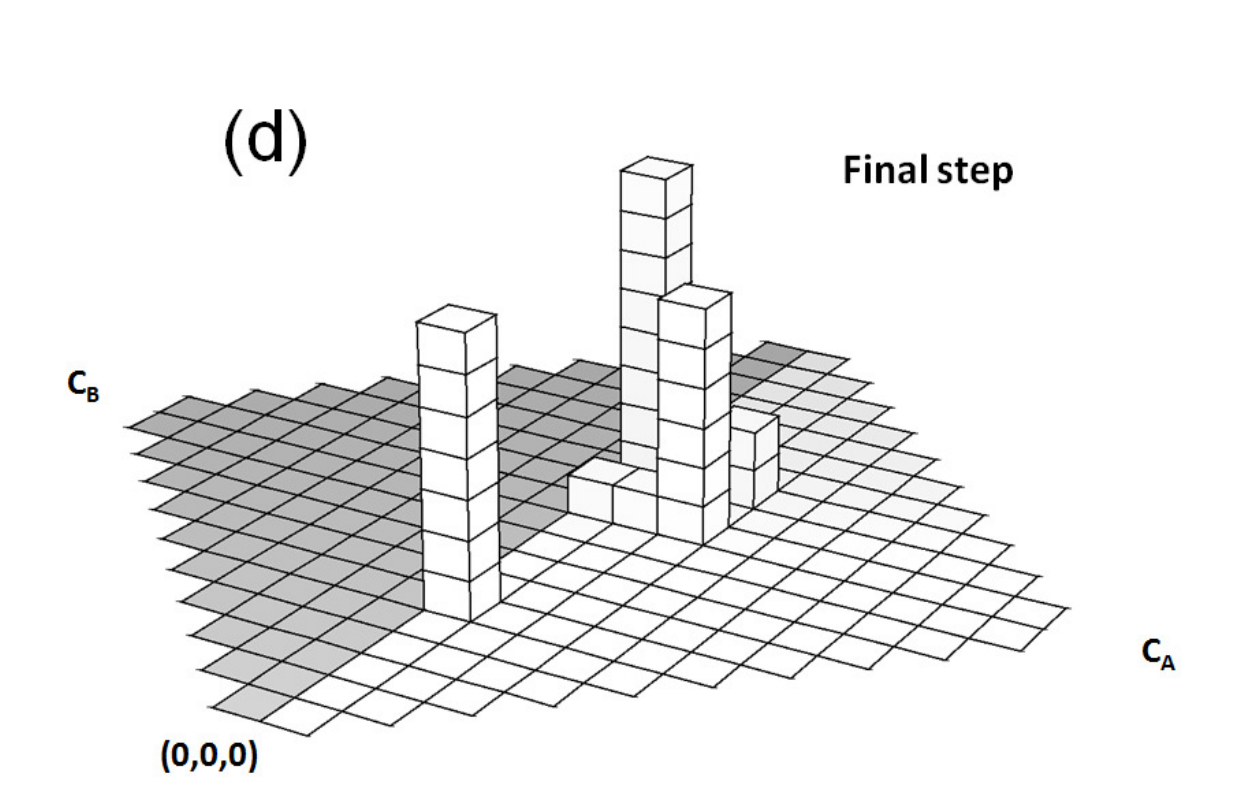}
\caption{Formation of consensus 4:(a) Plot of revealed preference $c_{A,B}$ versus number of steps $T$ 
with $\alpha=0.5$, $\mu=0.5$, (b)  frequency of $c_{A,B}$ at the first step, (c)  frequency of $c_{A,B}$ at the $400^{th}$ step, (d)  frequency of $c_{A,B}$ at the final step. Consensus reached is choice A.} \label{fig4}
\end{figure}

\begin{figure}[ht] \centering
\includegraphics*[width=8cm,height=8cm]{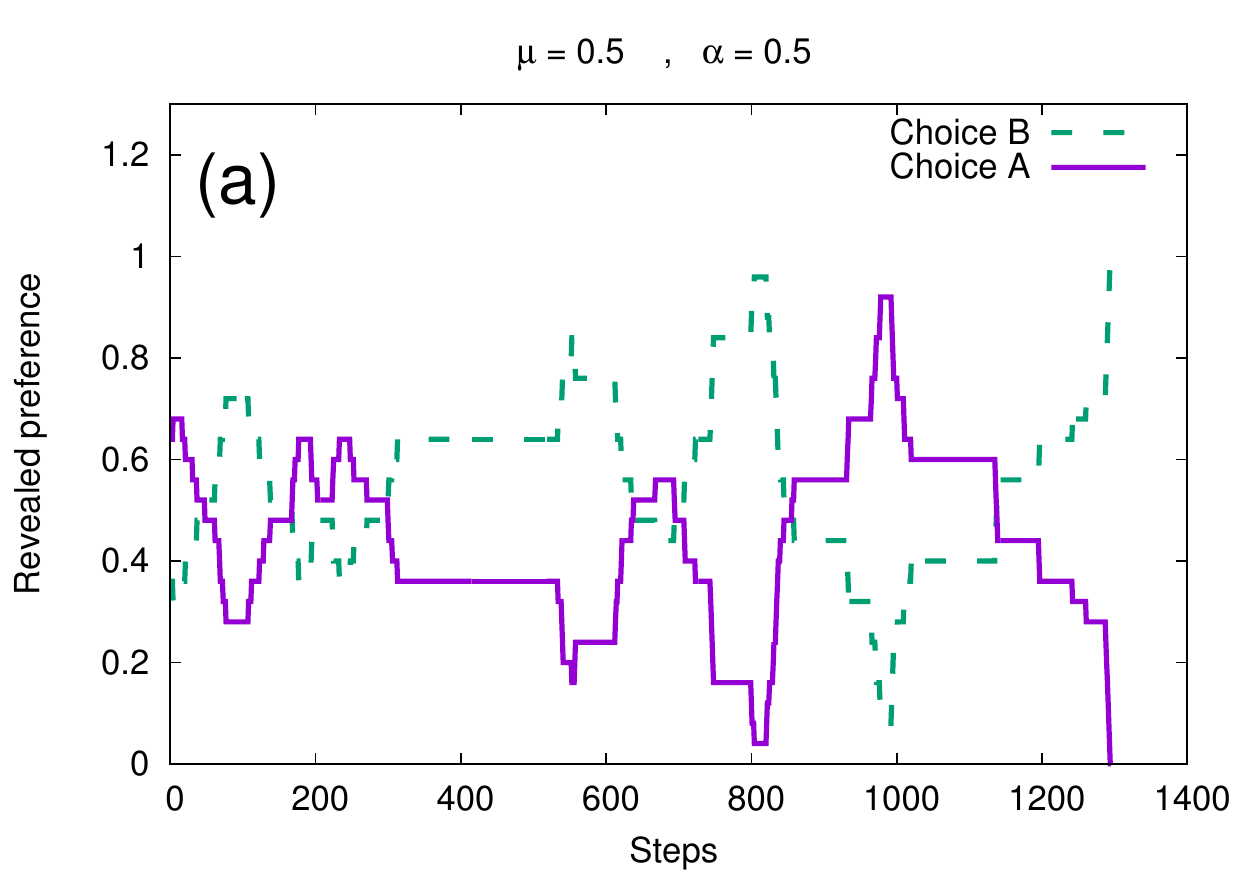}\hspace{1cm}
\includegraphics*[width=8cm,height=8cm]{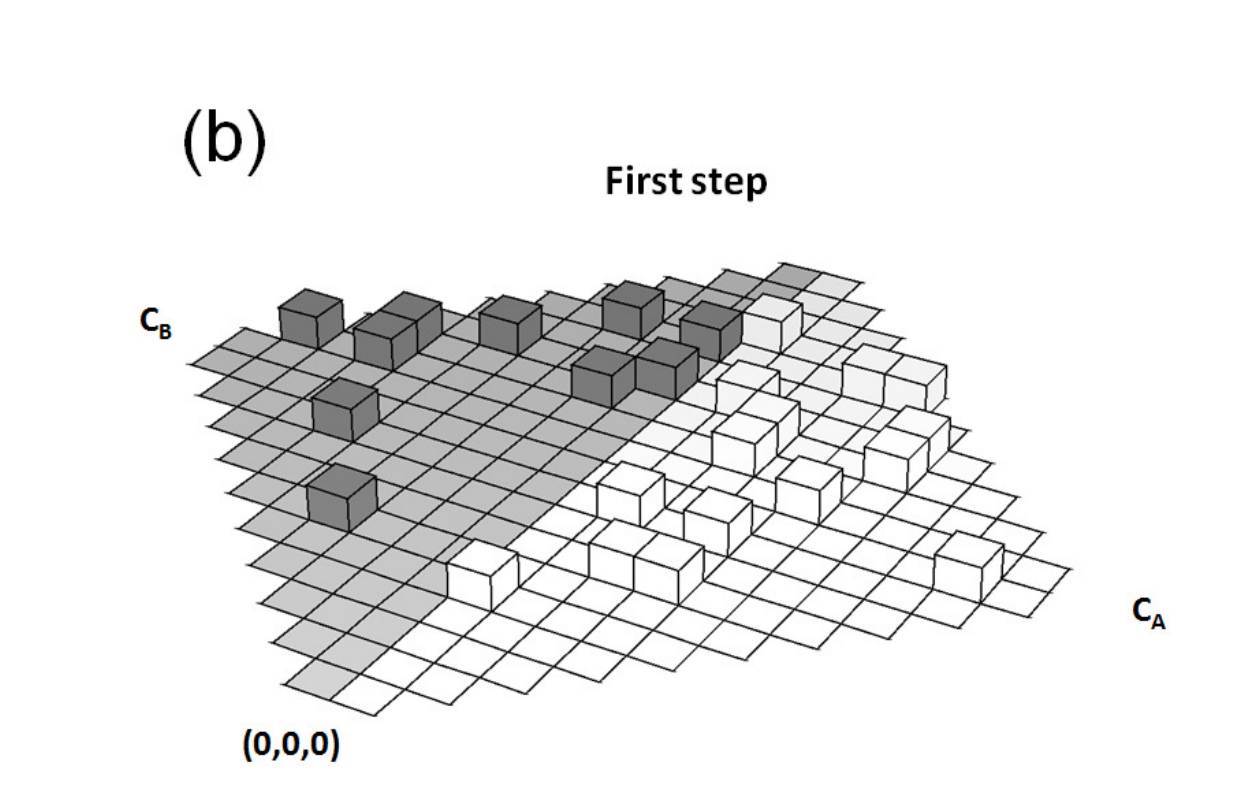}\\
\includegraphics*[width=8cm,height=8cm]{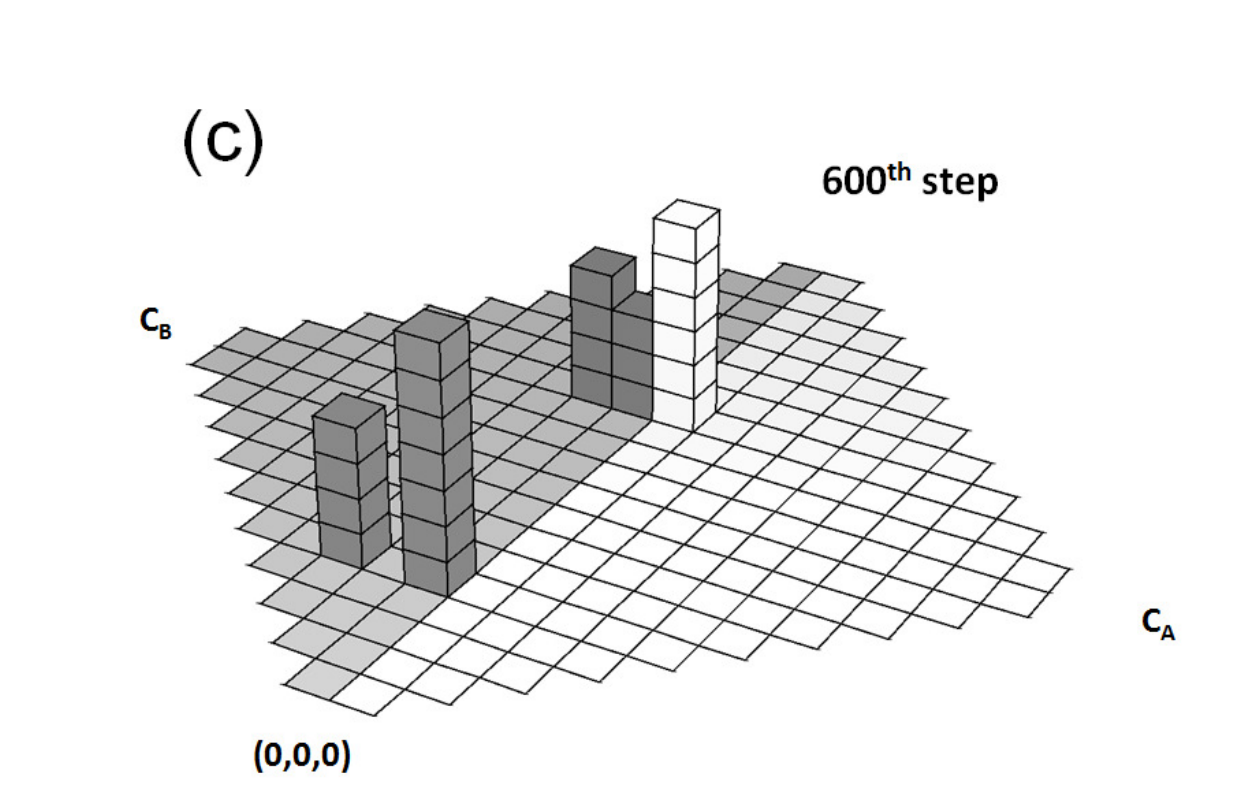}\hspace{1cm}
\includegraphics*[width=8cm,height=8cm]{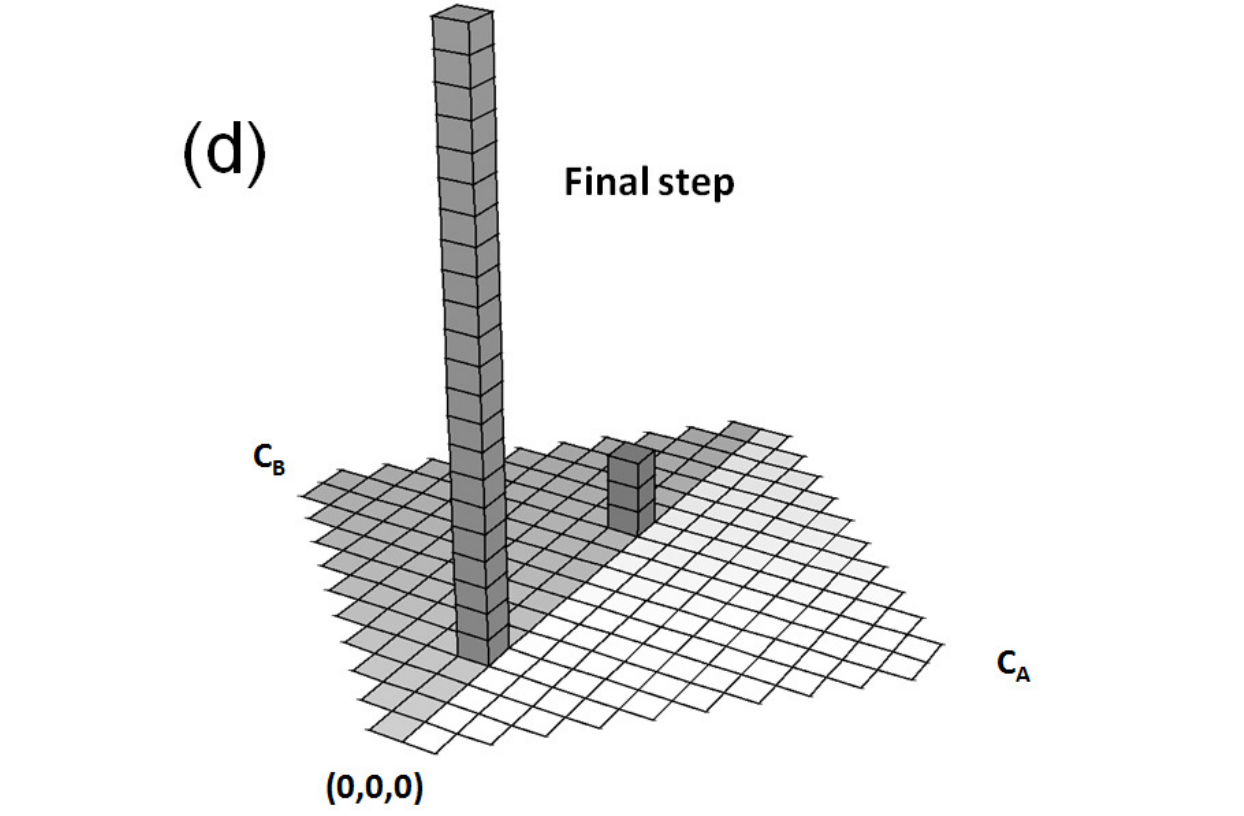}
\caption{Formation of consensus 5:(a) Plot of revealed preference $c_{A,B}$ versus number of steps $T$ 
with $\alpha=0.5$, $\mu=0.5$, (b)  frequency of $c_{A,B}$ at the first step, (c)  frequency of $c_{A,B}$ at the $600^{th}$ step, (d)  frequency of $c_{A,B}$ at the final step. Consensus reached is choice B.} \label{fig5}
\end{figure}


\begin{figure} 
 \centering
 \includegraphics{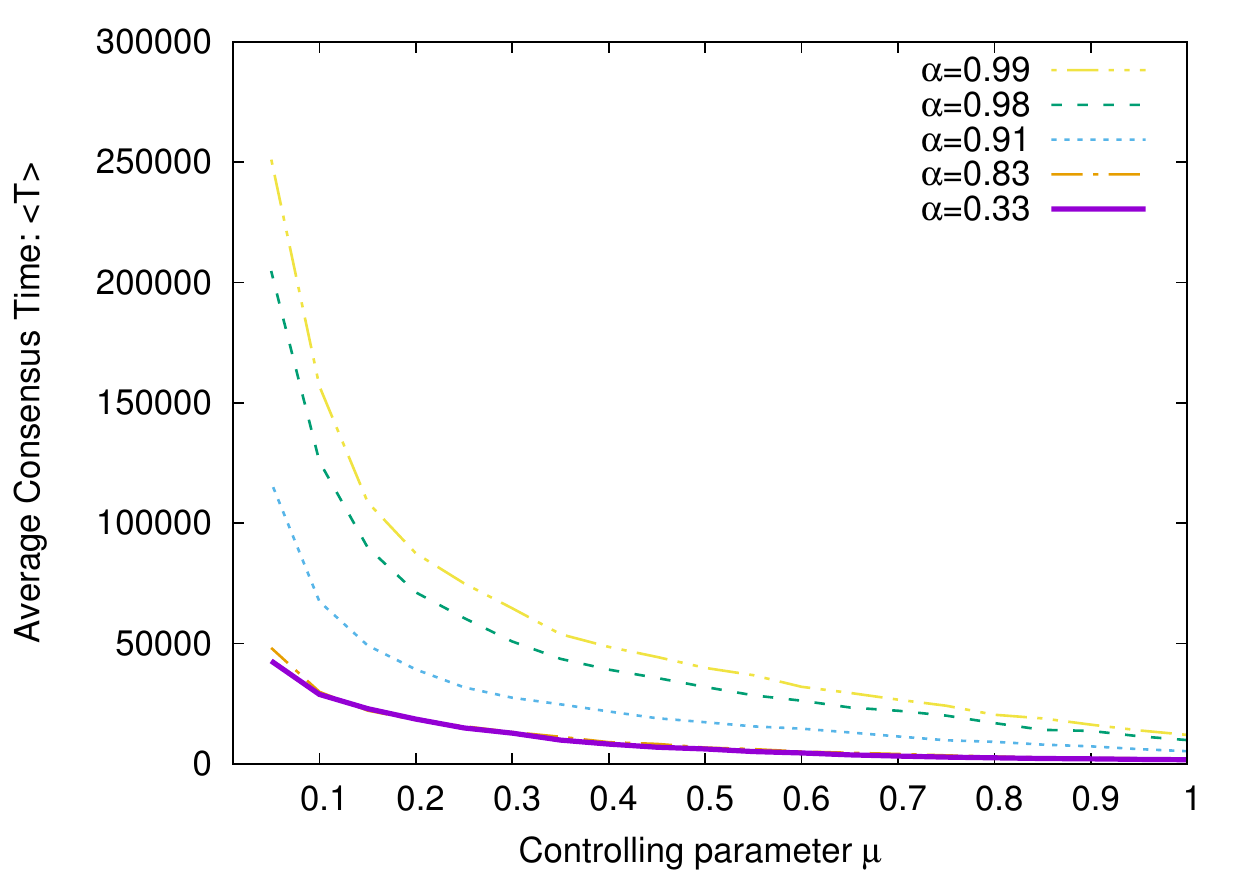}
\caption{The average consensus time $\langle T\rangle$ versus $\mu$ for different $\alpha$.}
\label{fig6}
\end{figure}

\begin{figure} 
 \centering
 \includegraphics{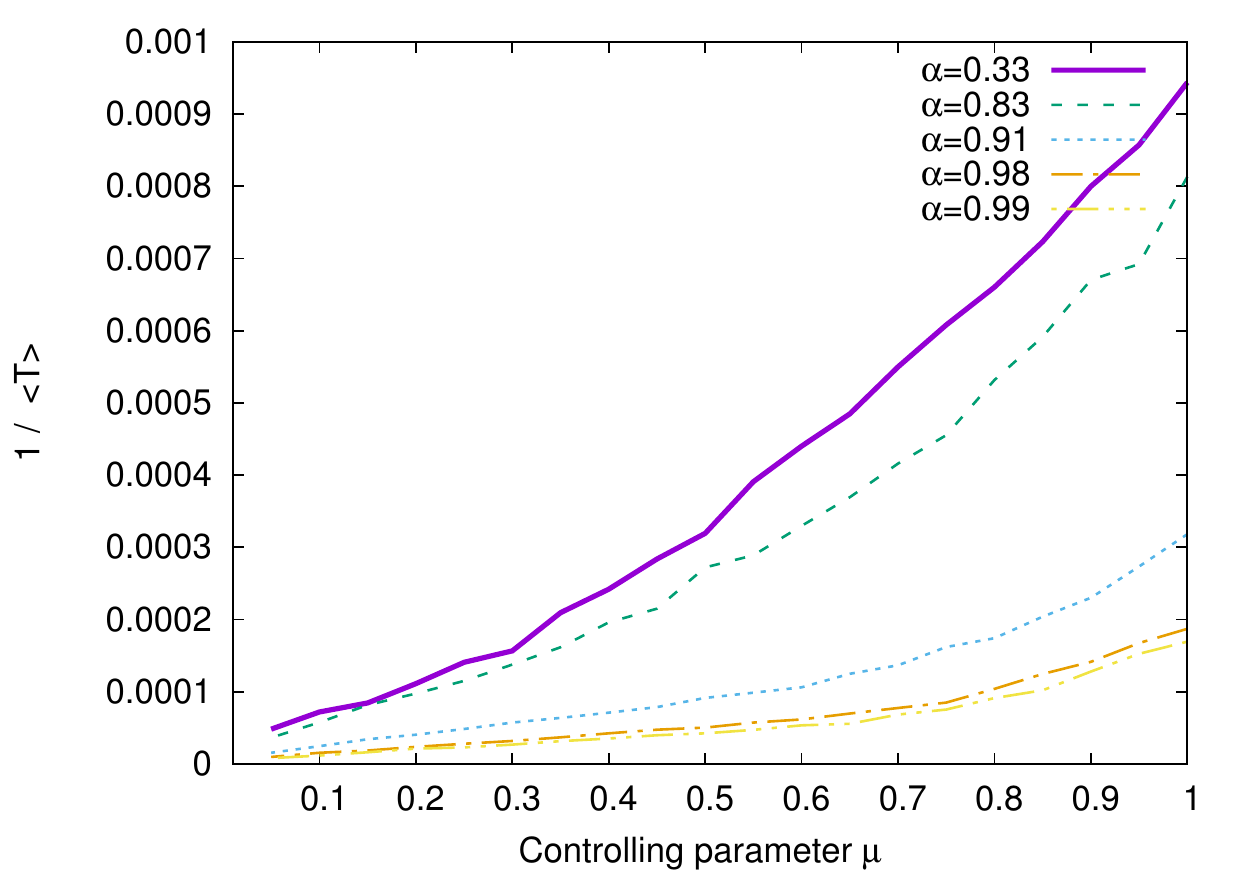}
\caption{The reciprocal of average consensus time  $1/\langle T\rangle$ versus $\mu$ for different $\alpha$.}
\label{fig7}
\end{figure}

\begin{figure} 
 \centering
 \includegraphics{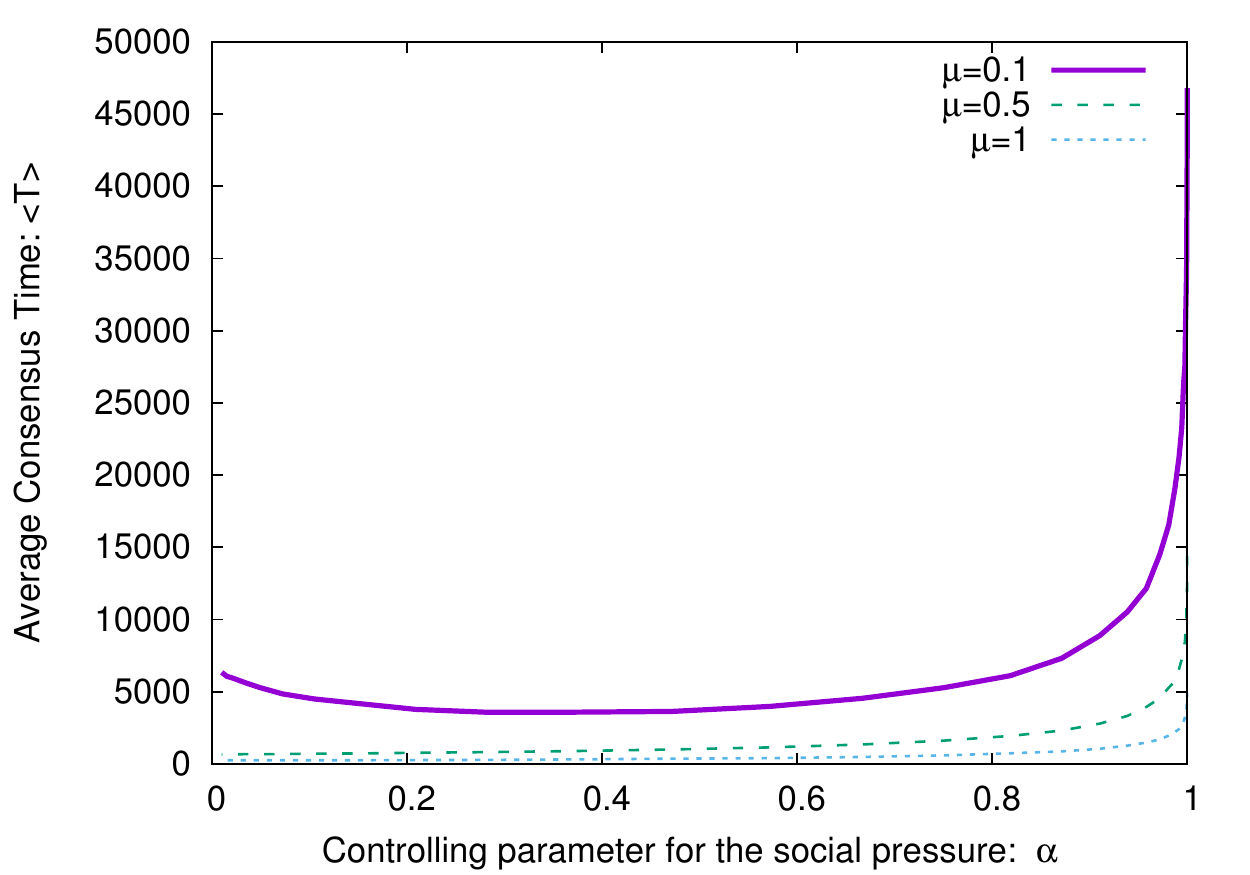}
\caption{The average consensus time $\langle T\rangle$ versus $\alpha$ for different $\mu$.}
\label{fig8}
\end{figure}

\begin{figure} 
 \centering
 \includegraphics{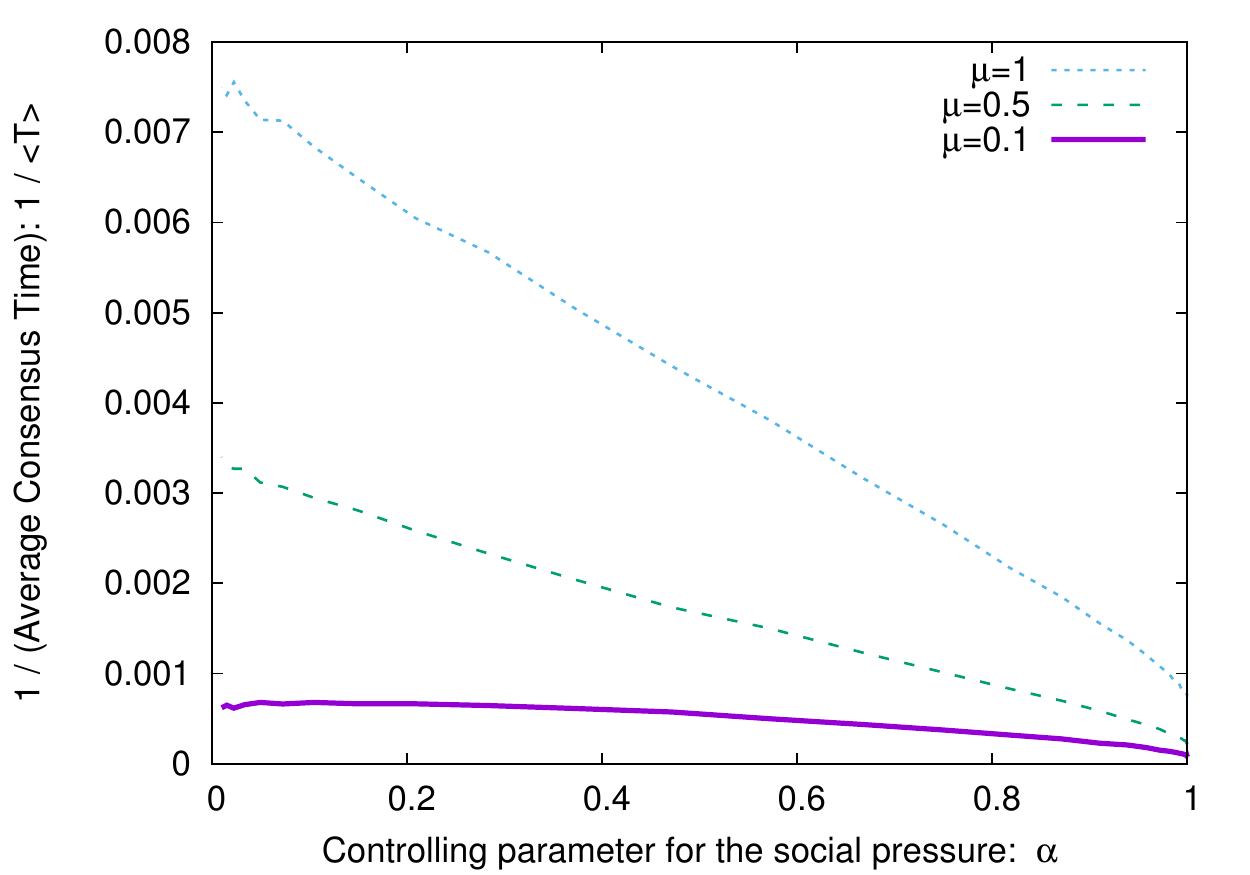}
\caption{The reciprocal of average consensus time $1/\langle T\rangle$ versus $\alpha$ for different $\mu$.}
\label{fig9}
\end{figure}

\begin{figure} 
 \centering
 \includegraphics{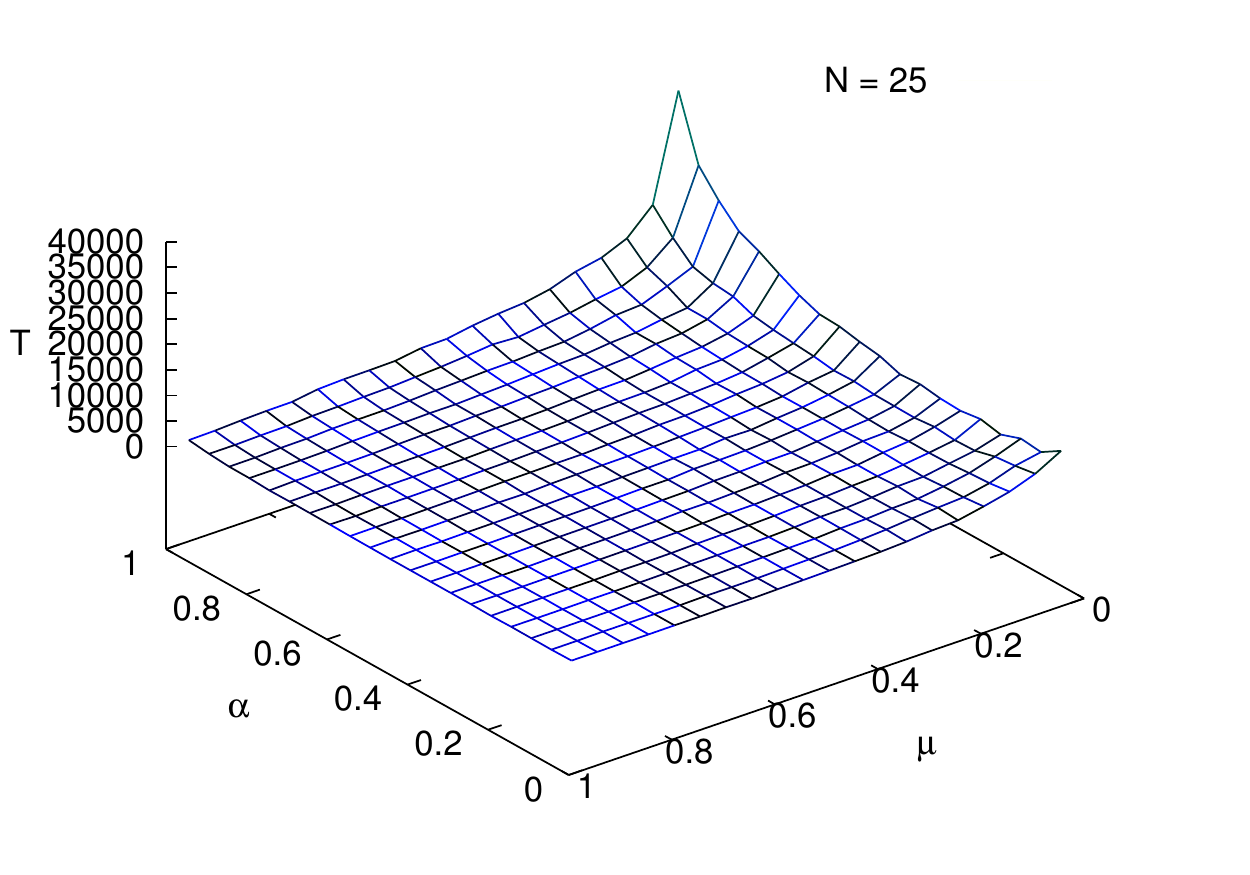}
\caption{The 3-dimensional plot of $\langle T\rangle$ versus $\alpha$ and $\mu$ for $N=25$ agents with an ensemble size of $200$ for each set of parameters.}
\label{fig10}
\end{figure}

\begin{figure} 
 \centering
 \includegraphics*[width=16cm]{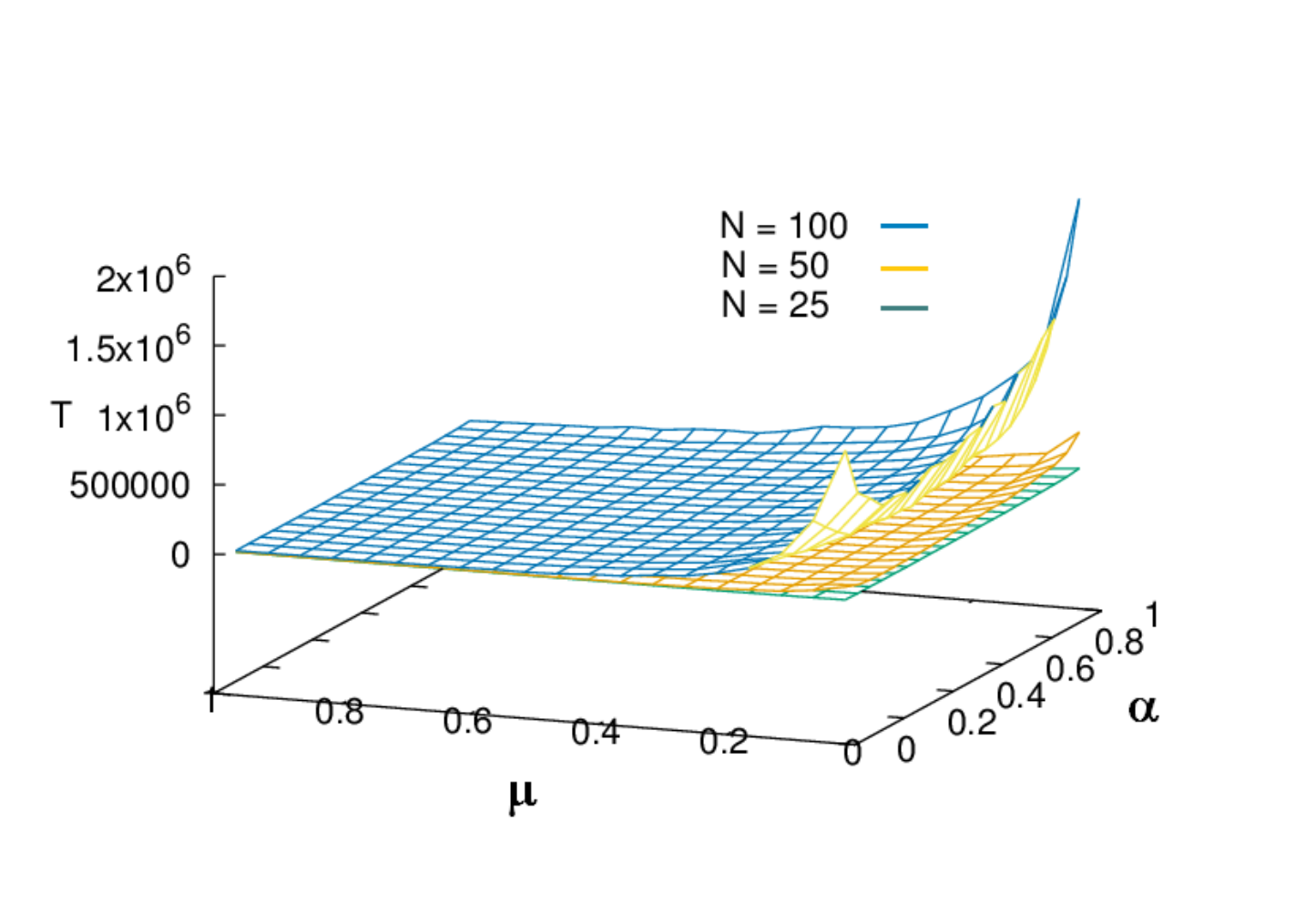}
\caption{The 3-dimensional plot of $\langle T\rangle$ versus $\alpha$ and $\mu$ for $N=25, 50$ and $100$ agents with an ensemble size of $200$ for each set of parameters.}
\label{fig11}
\end{figure}

\end{document}